# Roman CCS White Paper

# Adding Fields Hosting Globular Clusters To The Galactic Bulge Time Domain Survey

**Roman Core Community Survey:** *Galactic Bulge Time Domain Survey*

**Scientific Categories:** *exoplanets and exoplanet formation; stellar physics and stellar types; stellar populations and the interstellar medium*

**Additional scientific keywords:** *Chemical composition, transits, variable stars, globular star clusters*

**Submitting Author:**
Name: Samuel K. Grunblatt
Affiliation: Johns Hopkins University
Email: sgrunbl2@jhu.edu

**List of contributing authors** (including affiliation and email):
Robert F. Wilson, NASA Goddard, robert.f.wilson@nasa.gov
Andrew Winter, Observatoire de la Côte d'Azur, andrew.winter@oca.eu
B. Scott Gaudi, Ohio State University, bsgaudi@osu.edu
Daniel Huber, University of Hawaii/ University of Sydney, huberd@hawaii.edu
Daniel A. Yahalomi, Columbia University, daniel.yahalomi@columbia.edu
Andrea Bellini, STScI, bellini@stsci.edu
Zachary R. Claytor, University of Florida, zclaytor@ufl.edu
Jorge Martinez Palomera, BAERI, palomera@baeri.org
Thomas Barclay, NASA Goddard, thomas.barclay@nasa.gov
Guangwei Fu, Johns Hopkins University, guangweifu@gmail.com
Adrian Price-Whelan, Flatiron Institute, aprice-whelan@flatironinstitute.org

**Abstract:** *Despite multiple previous searches, no transiting planets have yet been identified within a globular cluster. This is believed to be due to a combination of factors: the low metallicities of most globular clusters suggests that there is significantly less planet-forming material per star in most globular clusters relative to the solar neighborhood, the high likelihood of dynamical interactions can also disrupt planetary orbits, and the data available for globular clusters is limited. However, transiting planets have been identified in open clusters, indicating that there may be planets in more massive clusters that have simply gone undetected, or that more massive clusters inhibit planet formation. Less than two degrees away from the nominal Galactic Bulge Time Domain Survey footprint, two globular clusters, NGC 6522 and NGC 6528, can be simultaneously observed by the Roman telescope during the Galactic Bulge Time Domain Survey. These clusters are comparable in mass (1-2 x $10^5$ solar masses) and age (~12 Gyr), but feature drastically different average metallicities— NGC 6522 has an average [Fe/H] ~ -1.3, while NGC 6528 has an average [Fe/H] ~ -0.1. If no transiting planets are detected in one season of time domain observations of these clusters, this would indicate a difference in planet occurrence among field stars and globular clusters at >3-σ significance even after accounting for metallicity, which could be enhanced to >5-σ significance with similar observations of another nearby field hosting a metal-rich globular cluster. This will reveal whether metallicity plays the dominant role in planet formation suppression in globular clusters. Additionally, time domain observations of NGC 6522 and NGC 6528 will detect variable stars in both clusters, testing the connection between stellar variability and binary fraction to metallicity and cluster environment, as well as testing the dependence of exoplanet yields on stellar density and distance from the Galactic midplane.*

**The stellar populations of globular clusters are not well understood.** Globular clusters tend to be very old stellar populations, but have a wide range of compositions, from solar-like to incredibly metal-poor. In addition, multiple populations of stars with unique abundance patterns have been identified in many globular clusters that do not follow the trends seen in the Galactic disk, and these trends have unique additional features within individual globular clusters (Bastian & Lardo 2018, Gratton et al. 2019). Finally, the abundance trends seen in globular clusters are not echoed in open clusters, which tend to be young and less massive. More detailed studies of the stars in globular clusters would assist with understanding these unique environments.

**Do globular cluster environments suppress formation of planets?** One of the most intriguing details of globular cluster populations is the apparent lack of planets. Despite a continuous 8.3-day survey of over 34,000 stars in 47 Tucanae with the Hubble Space Telescope, no transiting planets were detected (Gilliland et al. 2000). This result was confirmed further through additional followup of this and other globular clusters (Weldrake et al. 2005, Weldrake et al. 2008, Nascimbeni et al. 2012, Wallace et al. 2020). This lack of planet detection has been suggested to have many root causes—inhibition of planet formation by massive star irradiation (e.g. Armitage 2000, Thompson 2013), loss of planets during close stellar encounters over time (Spurzem et al. 2009, Winter & Clarke 2023), or low average metallicities of globular clusters have all been suggested. Evidence for the correlation between planet occurrence and metallicity has been found repeatedly (Fischer & Valenti 2005, Johnson et al. 2010, Petigura et al. 2018), but the effect of enhanced stellar density is hard to test outside of the extreme environments of globular clusters. Thus, the independent effects of metallicity and cluster environment have not been isolated, and thus it is not clear what is dominating the apparent suppression of planet occurrence in globular clusters. Furthermore, the significance of the non-detection of planets in globular clusters may be lower than initially anticipated.

Masuda and Winn (2017) re-analyzed the planet yield of the first survey of Gilliland et al. (2000), using the transiting planet yields of the Kepler survey. This study finds that despite the continuous observation of 34,000 stars in the core of 47 Tuc over 8.3 days, the expected yield can now be estimated to be 2 +/- 1 planets, a factor of ~8 smaller than was initially predicted by Gilliland. This original overestimate was due to an overestimate of the occurrence rate and average radius of transiting hot Jupiters extrapolated from the sole known transiting hot Jupiter at the time, HD209458b (Charbonneau et al. 2000, Henry et al. 2000). The updated yield is derived using updated occurrence rates from Kepler (Howard et al. 2012, Thompson et al. 2018) and accounts for the lower average mass of main sequence stars present in globular clusters (Bryant et al. 2023), but does not account for stellar metallicity. The planet-metallicity correlation also suggests that giant planet occurrence $f$ is proportional to $10^{1.2*[Fe/H]}$, and thus is significantly lower around metal-poor stars. As the average metallicity of 47 Tuc is -0.8 dex lower than solar, this brings the expected planet yield below unity for the Gilliland et al. (2000) survey. Thus, it is unclear what role the unique dynamical environment of a globular cluster plays with regards to planet occurrence.

**Roman can probe globular clusters for planets more efficiently than any other previous mission.**
The Nancy Grace Roman Space Telescope is a uniquely valuable tool for studying globular clusters. The high resolution yet extremely wide field of view means that essentially all stars in a given globular cluster can be imaged by the Roman telescope in a single pointing. Furthermore, the astrometric precision of Roman implies that relatively short baselines are needed to get precise membership lists for almost all globular clusters (Bellini et al. 2019). These kinematic membership determinations will also explore effects of the Galactic tidal field, and reveal tidal tails of globular clusters. Finally, the use of time domain observations of clusters would allow transiting planet detection in globular clusters, as has been shown to be possible in open clusters observed by Kepler (e.g., Meibom et al. 2013). Given the smaller pixels and redder passband compared to Kepler, and the much larger field of view compared to HST or JWST, Roman is optimally designed to produce a much larger yield of light curves for a given star cluster than any currently operating facility.

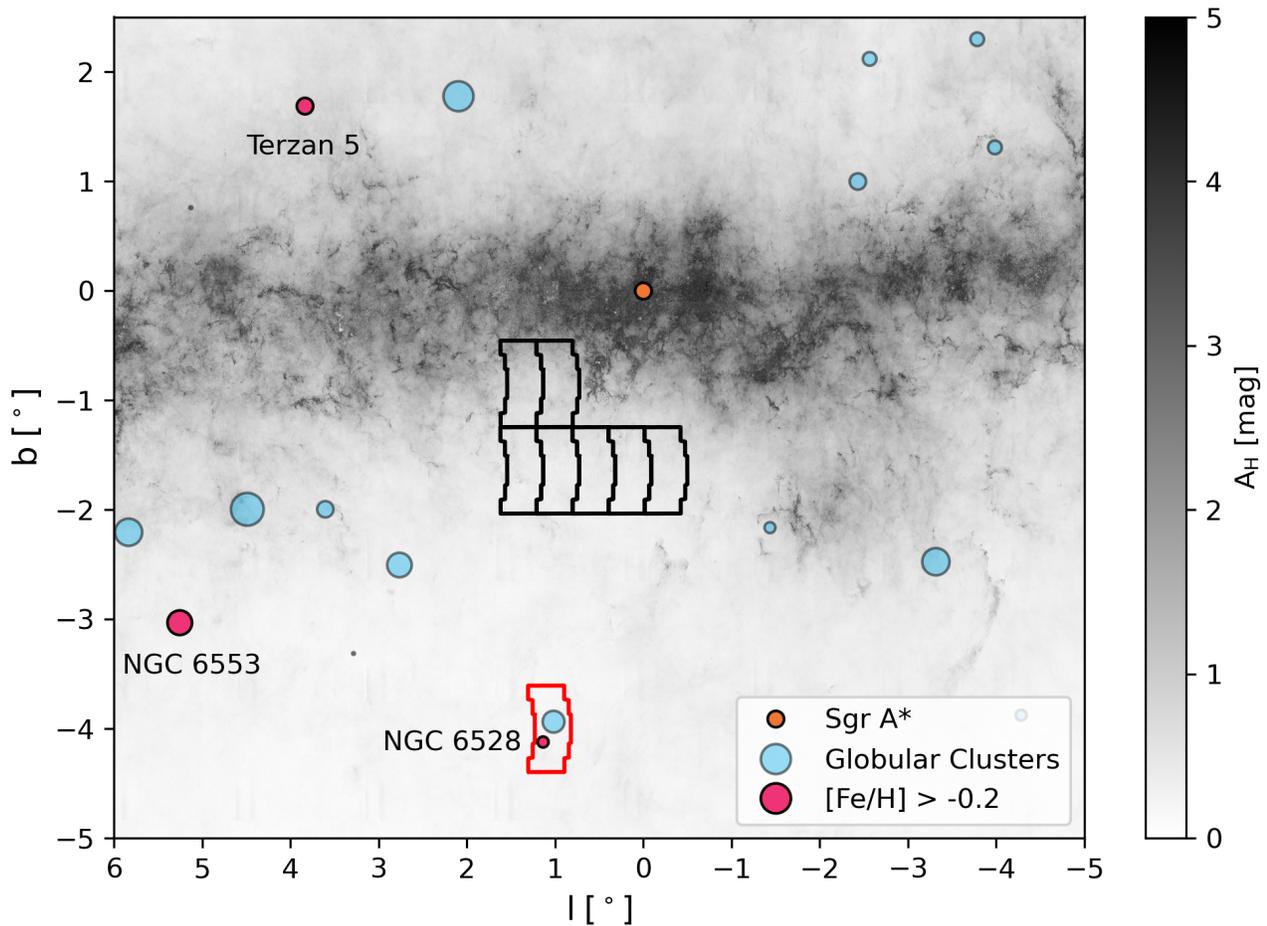

Figure 1: The footprint of the proposed Galactic Bulge Time Domain Survey, overlaid on a dust map showing magnitudes of distinction in the H band of the Galactic Center (Surot et al. 2020). We have designated globular clusters as circles on this plot, where the point size is relative to the angular size of the cluster. Solar-like metallicity globular clusters have been highlighted and labeled. We demonstrate that both NGC 6522 and NGC 6528 can be observed in one Roman pointing only a couple of degrees away (red footprint) from the bulge survey fields.

**Extensions to the Roman Galactic Bulge survey would allow for comparative globular cluster studies.** The observing plan for the Roman Galactic Bulge Time Domain Survey (GBTDS) has left room for additional time-domain observations in nearby fields at nearly the cadence of the main survey. Here we explain how this could be used for comparative globular cluster studies, and could distinguish the causes and effects of different planet formation suppression mechanisms in globular clusters. Figure 1 illustrates the nominal galactic bulge survey footprint, overlaid on a dust map of the Galactic Center, illustrating magnitudes of extinction in the H band, a similar passband to the widest bandpass available with Roman (Surot et al., 2020). We have identified another field approximately 2 degrees south of the footprint, which hosts two globular clusters, NGC 6528 and NGC 6522. These clusters are close in mass (1-2 x $10^5$ $M_{Sun}$) and age (12 Gyr) but have very different metallicities—NGC 6528 has a mean metallicity close to solar ([Fe/H] = -0.14+/- 0.03, Muñoz et al. 2018), while NGC 6522 has a more typical globular cluster metallicity ([Fe/H] = -1.0 +/- 0.2, Schiavon et al. 2017). With one additional pointing, Roman can capture both clusters and their surroundings. At a cadence of 30-minutes or higher over one observing season, this should be sensitive to stellar variability, as well as transiting/eclipsing events with periods of 25 days or less.

**Comparison to the HST transit study of 47 Tuc implies that no detections of transiting planets in this survey would be highly significant.** Previous ground-based photometric studies with a similar survey area in a similar bandpass, with a pixel scale 3 times larger and a limiting magnitude 3 magnitudes

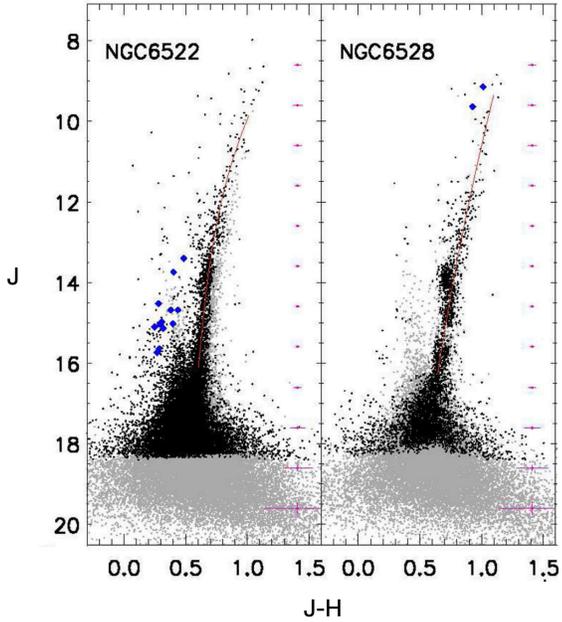

Figure 2: Color-magnitude diagrams from VVV photometry of NGC 6522 and NGC 6528. Stars passing their contamination criteria are shown in black. Photometric errors are shown in pink, and stars with spectroscopic data are shown as blue diamonds. Asteroseismology should be possible for stars at or brighter than the red clump (J~14). Adopted from Cohen et al. (2017).

fainter than Roman, identified 4,200 cluster stars with precise photometry in NGC 6528 and 15,000 such stars in NGC 6522 (Cohen et al., 2017, see Figure 2). As stellar mass function studies imply the number of stars in globular clusters should increase more strongly than linearly with magnitude over the range of magnitudes that Roman is sensitive to in bulge globular clusters (Baumgardt et al., 2023), and the number of sources we can precisely resolve in a cluster should increase roughly linearly with pixel scale (Coe et al., 2016), we predict that we should be able to produce light curves of roughly 30,000 stars in NGC 6528 and 100,000 stars in NGC 6522 down to a magnitude $m_{F146} = 21.5$. We note that the photometric uncertainty predicted for an $m_{F146} = 21.5$ star observed by Roman over the course of one hour is approximately 1%, a factor of 3 more precise than the photometric uncertainty reached for an $m_V = 21.5$ star observed by the Gilliland et al. (2000) survey over the entire 8.3-day duration of the mission (Wilson et al. 2023, see Figure 3). This ratio holds for essentially all stellar magnitudes that will be covered in this survey. In addition, the length of one Roman observing season is more than 9 times that of the Gilliland et al. (2000) survey. Equation 1 of Howard et al. (2012) states that the signal to noise of a transiting planet signal is proportional to the square root of the number of transits (and therefore the baseline of observations). Thus, the signal to noise of a given planet transit in our proposed observations should be at least 9 times larger in this survey than it would have been in the Gilliland et al. (2000) survey for a star where $m_V = m_{F146}$, due to the >3x higher photometric precision and 9x longer baseline. As the majority of the main sequence stars in globular clusters are less massive and therefore redder than the Sun, $m_V > m_{F146}$ for most targets, thus implying a slightly larger signal to noise for a given target. Furthermore, planet occurrence increases exponentially for smaller planet radii, and is uniform with the logarithm of orbital period (Howard et al. 2012), and thus we expect the planet yield to increase more than the relative signal to noise increase for a given target in our survey. As the Gilliland et al. (2000) survey was signal to noise limited, and the expected planet yield determined by Masuda & Winn (2017) assuming a solar metallicity was 2 +/- 1 transiting planets among 34,000 stars, we naively expect to detect ~70 planets in our survey before accounting for metallicity. *After taking the planet-metallicity correlation and uncertainty in well-resolved star counts into account, we expect to detect 13 +/- 4 transiting planets in NGC 6528 and ~1 planet in NGC 6522. This implies the detection of no transiting planets would be inconsistent with planet occurrence observed by Kepler and TESS at >3-σ significance.* We note that this detection rate also agrees with observations of the Galactic Bulge with HST, which demonstrated that 16 transiting planet candidates and at least 2 confirmed planets could be detected among tens of thousands of bulge stars observed with worse photometric precision and a lower cadence, but also a smaller pixel scale, than what Roman can achieve for bulge globular clusters (Sahu et al., 2006).

The simultaneous observations of a similar or larger number of stars in NGC 6522 will further test the rate of planet occurrence in clusters, and if transiting planets are detected in NGC 6528 but not in NGC 6522, this suggests the planet-metallicity correlation dominates cluster planet occurrence trends. *Additionally, NGC 6553, located only 3 degrees further from the proposed bulge fields than NGC 6528, has been shown to have almost 6 times the number of well-resolved stars, yet a similar composition to NGC 6528 (Cohen et al. 2017, Muñoz et al. 2018), suggesting that one season of observing this cluster*

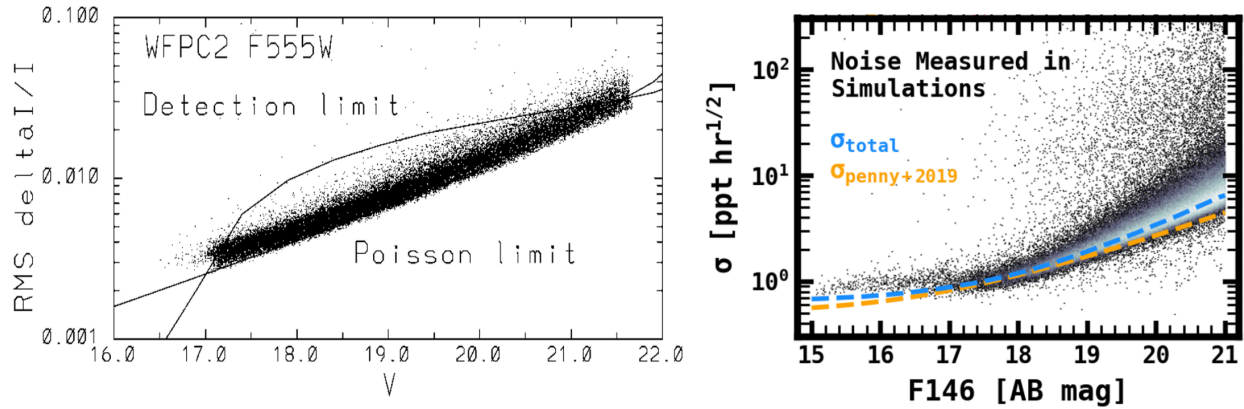

Figure 3: Left: RMS scatter over the 8.3-day observation baseline per target vs. V magnitude observed by HST/WFPC2 as part of the Gilliland et al. (2000) survey of 47 Tuc. Taken from Gilliland et al. (2000). Right: RMS scatter over one hour per target vs. F146 magnitude simulated from the Roman GBTDS. Taken from Wilson et al. (2023). Roman scatter should be roughly 25% or less than that of WFPC2 targets at a given magnitude, and since globular cluster main sequence stars are lower-mass, they will be multiple magnitudes brighter in F146 than in the F555 (V) band pass.

***with 30- to 60-minute cadence either instead of or along with NGC 6522 and NGC 6528 could test the comparative occurrence rate between cluster environments and the Kepler field at >5-σ significance.*** Similar observations of a third nearby metal-rich cluster, Terzan 5, could further probe the dependence of planet occurrence on cluster environmental factors (see Figure 1).

**Serendipitous detections in these fields will probe Galactic bulge structure.** Beyond planet occurrence in globular clusters, these observations will inform other studies. Multi-season observations of these fields with cadences of 30-minutes or better will also be sensitive to microlensing events. The rate of microlensing events is expected to be a factor of 3 lower in the vicinity of NGC 6528 as opposed to the proposed bulge fields (Sumi & Penny 2016, Penny et al. 2019), but even in only one season of observation, 10 events are expected. The rate of these events is determined by the column density of stars and thus the bulge structure, so measuring this event rate as a function of galactic latitude will produce better estimates of the Galactic bulge structure. Observations of NGC 6553 and Terzan 5, with absolute galactic latitudes smaller than that of NGC 6528, will also result in more microlensing detections and thus test bulge potential models. Furthermore, the number distribution of field stars, such as red giants, solar-like stars, M dwarfs, and white dwarfs, in these fields will be large enough to directly probe the star and transiting planet population distributions for these stars, potentially revealing hundreds to thousands of additional transiting planets and the structure of the Galaxy in the direction of the Galactic bulge (Fantin et al., 2020, Wilson et al., 2023, Tamburo et al., 2023).

**This study will also allow precise stellar characterization of tens of thousands of globular cluster stars.** Extrapolating from the precise stellar photometry yields of earlier surveys of bulge globular clusters, asteroseismology should be possible for essentially all red giants in the red clump or brighter in each globular cluster with time-series observations by Roman, a sample of hundreds to thousands of stars per cluster (Cohen et al. 2017, see Figure 2). Asteroseismology will allow precise determination of stellar masses and the stellar mass distribution of globular clusters, and potentially reveal compositional differences that are difficult to probe through other methods, such as stellar helium abundance (e.g. Verma et al., 2019). Stellar variability in globular clusters will also be captured, making it possible to determine flare and rotation rates and spot modulation amplitudes in thousands to tens of thousands of stars, which have yet only been measured in younger clusters (e.g. Feinstein et al., 2020, Gaidos et al., 2023) and field stars (Santos et al., 2021, Howard et al., 2022). This will test potential differences in stellar activity cycles and probe the evolution of stellar rotation at old ages, providing the first large sample of old anchors for gyrochronology relations. This study will also probe the rate of stellar binarity in globular clusters, which has been shown to be related to multiple populations and cluster metallicities (d'Orazi et al., 2017). Finally, cluster membership will be easily determined through proper motion detections of cluster stars, which should be achievable with at most two seasons of observations (Bellini et al., 2019).

**Transiting planet demographics in globular clusters can lead to extragalactic planet demographics.** Beyond the Roman GBTDS, the successful constraint of planet occurrence and detection of planets in globular clusters will motivate studies for planets in other clusters not associated with the bulge, and with potentially more exotic origins and stellar populations. For example, observations have revealed the existence of multiple stellar populations in the proposed nuclear star cluster of the Sagittarius dwarf galaxy, M54 (Carretta et al., 2010). The most metal-rich stars in M54 have metallicities comparable to NGC 6528 and do not have the same spatial distribution as the rest of the cluster (Kacharov et al., 2022). Time domain observations of this metal-rich population, or other metal-rich globular clusters associated with Sagittarius such as Terzan 7, with facilities such as Roman or JWST could result in the first independently verifiable detection of a transiting planet with an extragalactic origin.